\documentclass[aps,prb,twocolumn,groupedaddress,showpacs]{revtex4}

\draft % marks overfull lines with a black rule on the right
\usepackage{graphicx}% Include figure files
\usepackage{dcolumn}% Align table columns on decimal point
\usepackage{bm}

\begin{document}

%\preprint{}

\title{Modeling the underlying mechanisms for organic memory devices: Tunneling, electron emission and oxygen adsorbing}

\author{Yao Yao}
\email[Electronic mail: ] {yaoyao@fudan.edu.cn} \affiliation{State
Key Laboratory of Surface Physics and Department of Physics, Fudan
University, Shanghai 200433, China}
\author{Yintao You}\affiliation{State Key Laboratory of Surface Physics and Department of Physics, Fudan University, Shanghai 200433, China}
\author{Wei Si}\affiliation{State Key Laboratory of Surface Physics and Department of Physics, Fudan University, Shanghai 200433, China}
\author{Chang-Qin Wu}\affiliation{State Key Laboratory of Surface Physics and Department of Physics,
Fudan University, Shanghai 200433, China}

\date{\today}
\begin{abstract}
We present a combined experimental and theoretical study to get
insight into both memory and negative differential resistance (NDR)
effect in organic memory devices. The theoretical model we propose
is simply a one-dimensional metallic island array embedding within
two electrodes. We use scattering operator method to evaluate the
tunneling current among the electrode and islands to establish the
basic bistable I-V curves for several devices. The theoretical
results match the experiments very well, and both memory and NDR
effect could be understood comprehensively. The experimental
correspondence, say, the experiment of changing the pressure of
oxygen, is addressed as well.
\end{abstract}


\maketitle

Memory devices based on metal-insulator-metal resistive structure
with organic materials are promising in application.\cite{Review}
The underlying mechanism in these devices is not very clear up to
now due to the complicated phenomena observed, say for example, the
memory effect and the negative differential resistance (NDR) in ON
state, making an obstacle for controllable improvement of the
performance.\cite{Scott1,Coulomb} Two different but correlated
mechanisms have been addressed. The first one is the
Simmons-Verderber (SV) model based on the space charge accumulated
on some penetrated metallic islands.\cite{SV} This model could be
proposed to explain NDR, which is also closely related to the memory
effect.\cite{Scott1} In addition to it, electron tunneling and
Coulomb blockade were also proposed,\cite{Coulomb} indicating that
the different spatial distribution of the space charge gave rise to
the transition between ON and OFF state. The admittance spectroscopy
provided another evidence to the SV model.\cite{Admit} Whereas, the
basic drawback of this mechanism is that, the OFF state is always
observed, in practice, during the first scanning of bias voltage,
while SV model predicts the sample is initially in ON
state.\cite{Scott2} The second mechanism is the filamentary
conduction, stating that there would be some penetration of metal
atoms into organic layer during the deposition of top electrode,
which forms some conducting filaments under relative large bias
voltage, and the trap-controlled tunneling between different
filaments dominates the NDR effect.\cite{Fila1,Fila} The direct
experimental evidence for this mechanism is from the transmission
electron microscopy image, which shows the metallic filaments are
formed in ON state.\cite{Yang1} Meanwhile, the temperature
insensitivity of ON state current,\cite{Temperature} the time
dependent switch-on behavior,\cite{Hou1,You1,You2} and the scanning
voltage and conductance dependence\cite{Hou2} all go together to
support the filamentary mechanism. However, it is still hard to
imagine the destruction and recovery of an atomic filament could
respond so fast to the scanning voltage. Combining the above two
mechanisms, we can get a common idea that, during the deposition of
the electrode, the metallic atoms could go deeply into the organic
layer and form some islands in the film,\cite{Coulomb,Island} which
is crucial to the NDR effect no matter it originates from space
charge or filaments.

Another experiments were focusing on the role of interface. Some hot
spots have been found on the surface of the device, and these spots
exist only in ON state but disappear in OFF state.\cite{Temperature}
This in some sense implies the memory effect is most possibly caused
by the build-up of some interfacial metallic tips, i.e., the OFF
state from a flat metal-organic interface, while the ON state from
some tips formed by penetrating metallic atoms.\cite{Yang3}
Theoretical estimation shows that, at these tips the electric field
could be as large as $5\times10^6$V/cm, under which the field
emission of electrons works. In order to verify this statement, we
reported very recently an experiment on electric field and
temperature dependence of switching time of the devices from OFF to
ON state.\cite{You2} The conclusion there figured out the very
possibility that metallic tips at the interface matter in the memory
effect.

Combining the above two considerations, it seems appropriate to say
that, the NDR effect is from the metallic islands formed in the
organic layer, while memory effect is from the electron emission at
the interfacial metallic tips. Therefore, on the present stage, an
in-depth and quantitative theoretical work responding to the
relative experimental understanding is highly demanded. In this
Letter, we will first establish a theoretical model to simulate the
metallic islands and tips in the organic memory devices and then
calculate the tunneling current among the electrode and islands to
study the basic characteristic of different devices. Then the
experiment of changing the pressure of oxygen will be addressed to
verify the corresponding theory.

\begin{figure}
\includegraphics[angle=0,scale=0.4]{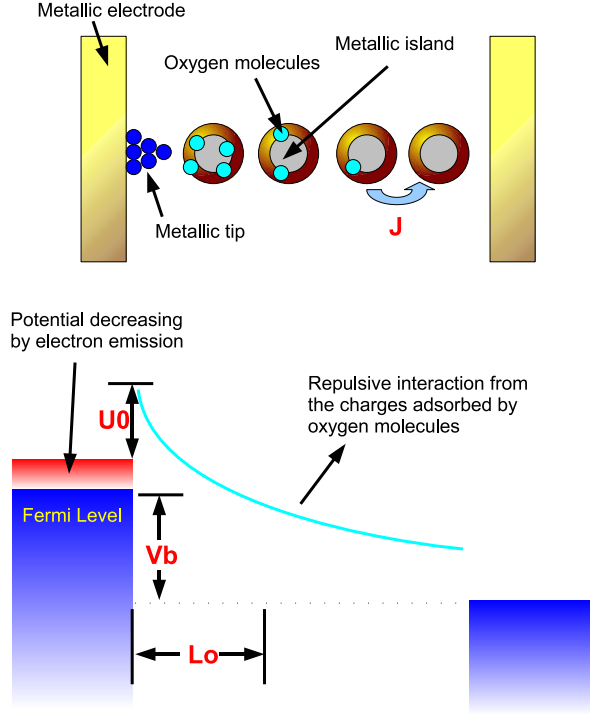}
\caption{Schematic of the structure of the system. The upper part is
for the device structure and the lower one is for the energy
diagram. See in the text for the detailed description.}\label{model}
\end{figure}

The model we consider in this work is briefly shown in Fig.
\ref{model}. The internal structure of the device is that, between
the two electrodes, there are some isolated metallic islands
surrounded by oxygen molecules. The islands are formed during the
deposition of top electrode, and the oxygen molecules are present
when we do the measurement of pressure dependence, as will be
discussed below. Because the islands are separated with electrodes,
the electrons injected from the electrode to the islands must first
tunnel the potential barrier $U_0$ formed by injection barrier of
normal organic molecules, which is around 1eV depending on the LUMO
or HOMO of organics and Fermi level of metals. The oxygen molecules
adsorbed by the metallic islands in the organic film will act as
deep traps to capture electrons, such that the barrier should be
largely increased due to the repulsion between the conducting
electrons and trapped electrons.\cite{Temperature} Here, for
simplicity, we have assumed that the distribution of adsorbed oxygen
is an exponential decay from the top electrode with the decay length
$L_o$, and the density of trapped electron is linear in space. The
transfer integral $J$ between neighboring islands is dependent on
the distance between these islands. Within this structure, the
metallic islands could form some conducting channels for electrons,
so that electrons could transport from one electrode to the other
via quantum tunnelings. The trapped electron in adsorbed oxygen
molecules will influence the potential profile of the islands, and
when the bias voltage is strengthened, the effect of NDR should be
found.\cite{Datta} On the other hand, when the device is switched to
ON state, some metallic tips at metal-organic interface will form as
discussed above, and then the injection barrier should be decreased
due to the electron emission. In this sense, the current becomes
larger and bistable effect emerges.

Based on the above description, we write down the model Hamiltonian
of the system as
\begin{eqnarray}
H=H_E+H_M+H_C.\label{hami}
\end{eqnarray}
The first term for the electrode is
\begin{eqnarray}
H_E=\sum_{\alpha,k}(\epsilon_{\alpha,k}\pm
\frac{V_b}{2})C^{\dag}_{\alpha,k}C_{\alpha,k},\label{hamiL}
\end{eqnarray}
where $\alpha=L/R$ is for left/right electrode,
$C^{\dag}_{\alpha,k}$ ($C_{\alpha,k}$) the electron creation
(annihilation) operator of the k-th continuum state in the
electrode, $\epsilon_{\alpha,k}$ the corresponding potential energy
of that state, $V_b$ the bias voltage, and $\pm$ taking $+$ for left
and $-$ for right. The second term of (\ref{hami}) for the metallic
islands is
\begin{eqnarray}
H_M=\sum_{i=1}^LJ(C^{\dag}_{i}C_{i+1}+{\rm
h.c.})+\sum_{i}U_0n^o_i\hat{n}_i,\label{hamiM}
\end{eqnarray}
where $C^{\dag}_{i}$ ($C_{i}$) creates (annihilates) an electron at
i-th metallic island, $\hat{n}_i=C^{\dag}_{i}C_{i}$, $J$ the
transfer integral between islands, and $n^o_i(\equiv f_iN_i)$ the
charges adsorbed by oxygen molecules, which plays the essential role
in this work. Wherein, $N_i [\sim \exp(-i/L_o)]$ the number of
oxygen molecule adhering to the metallic islands, $f_i$ the filling
factor, which will be assumed to be linear to $i$. The interfacial
injection potential induced by oxygen molecule is $U_0$, and the
length of the system is $L$, which is not crucial when it is changed
and will be set to $50$ throughout this work. In all, there are
three adjustable parameters in this theory: $U_0$ is the interfacial
injection potential, which determines the transition between OFF and
ON state; $J$ is the coupling between neighboring islands, which is
determined by how many islands are formed in the film, and we have
neglected the disorder of $J$; $L_o$ is the penetration length of
oxygen molecule, which is determined by the concentration of oxygens
and the trapped electrons in them as well. All these parameters
could be directly determined by the experiments, as is shown later.

The last term of (\ref{hami}) is the coupling between electrodes and
islands, which has the form
\begin{eqnarray}
H_C=\sum_{\alpha,k}J_{0}(C^{\dag}_{\alpha,k}C_{1/L}+{\rm
h.c.}),\label{hamiC}
\end{eqnarray}
where $J_0$ is the transfer integral between electrode and island.
This parameter is directly related to the spectral broadening
induced by electrodes, such that one can use it as a scaling to all
results and then compare them with the experimental measurements.

Under a given bias voltage, a nonequilibrium steady state (NESS)
could be achieved, and one can then calculate the corresponding I-V
curve. To evaluate the tunneling current of NESS, we utilize the
scattering operator method, which has been well developed by J. E.
Han for the system of quantum dots.\cite{Han} The basic idea of the
method is to derive the scattering operators from the
Lippmann-Schwinger equation and then evaluate the corresponding
tunneling current.\cite{Han,Hershfield} Basically, the method we use
in this work is straightforward from Han's treatment, such that we
would not go further to discuss the details of the method.

Based on the our model (\ref{hami}), we first try to rebuild the
basic experimental I-V curves of both ON and OFF states for
different device structures. The details of experiments are as
follows. All the materials used were commercially available without
any further purification. The devices were fabricated on ITO-coated
glass substrates with sheet resistance about $17\Omega$/sq. Each
substrate was heated to $150^{\circ}$C in air after solvent cleaning
and then loaded into a high vacuum chamber ($\sim2\times10^{-6}$Pa).
All the metallic and organic layers were formed by vacuum
depositions. The deposition rate of Alq$_3$, Mg and Al was about
$0.8{\rm \AA}$/s, $1{\rm \AA}$/s and $1{\rm \AA}$/s respectively.
The area of the organic memory device was 9mm$^2$, defined by
3-mm-wide ITO stripes overcrossed by 3-mm-wide top metal stripes.
The current-voltage (I-V) characteristics of the devices were
measured by a Keithley 236 unit with the top electrode grounded. The
measurements of electric properties of the devices were carried out
in a glove box (O$_2<$1ppm and H$_2$O$<$1ppm) or a dewar (Janis
Research Co. Inc. VTF-100) under 10$^{-1}$Pa pumped by a mechanical
pump.

\begin{figure}
\includegraphics[angle=0,scale=1.4]{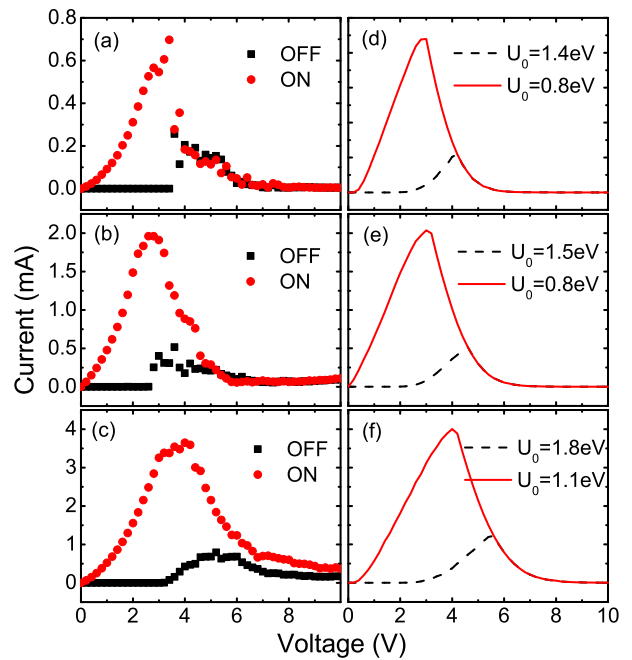}
\caption{Experimental (left panels) and corresponding theoretical
(right panels) results of both ON and OFF I-V curves for different
device structures: (a) ITO/Alq$_3$ (120nm)/Al; (b) ITO/Alq$_3$
(120nm)/LiF/Al; (c) ITO/Alq$_3$ (120nm)/Mg/Al; and different
parameters: (d) $J=0.35$eV, $L_o=0.1L$; (e) $J=0.4$eV, $L_o=0.1L$;
(f) $J=0.5$eV, $L_o=0.1L$.}\label{U0}
\end{figure}

In the left panels of Fig. \ref{U0}, we show experimental
measurements for three different structures by changing metallic
electrodes. One can find some typical characteristics for these
curves. For example, as the character of the memory effect, the OFF
state current has a turn-on point at around $3$V depending on the
specific structure, while the ON state current goes up very quickly
from 0V. Both ON and OFF state current have the similar line shape,
that is, a wavepacket-like shape in a certain region of voltage.
Especially, the downward sides of both states, say the NDR region,
overlap heavily. The ratio between ON and OFF state and also the
absolute value of them depend on the specific structure. In the
right panels of Fig. \ref{U0}, we show the corresponding theoretical
results. It is shown that, all the results are in good agreement
with the experiments. By changing $U_0$, we have obtained the
typical ON and OFF state current, respectively. That means, the
memory effect is shown to correspond to the interfacial potential
reduction, which is consistent with our previous experimental
results. \cite{You2} On the other hand, the NDR effect has also been
rebuilt by the theory, which is of importance in this type of memory
device as we have mentioned.

To explain the NDR effect we found, we summarize the whole physical
process inside our theory as follows: Initially, when the scanning
bias voltage starts from zero, all the energy potentials on each
metallic island are the same, and no current is present. Then the
voltage increases gradually, and the current follows to increase
instantaneously. At this moment, the potential on each island shifts
and starts to differ from each other. As we know, the largest
tunneling current emerges once the potential of each island
coincides. That means, under a certain voltage, depending on the
potential drop among the islands, the current will meet the maximum
value and then go down, so that NDR effect occurs.

It is worth noting that, for each curve of the theoretical results,
the $U_0$ change is about $0.7$V from OFF to ON. Thus, if the
metallic tip formed at the interface is estimated to be $10$nm, the
electric field it induces will be $7\times 10^5$V/cm, much smaller
than the estimation from experimenters.\cite{Yang1} The condition
for the bistable effect becomes much softer in our theory.
Meanwhile, the value of $U_0$ for each device at OFF state is
comparable to the injection barrier of electron, which is consistent
with the fact that the metallic islands, as donors, matter in the
whole process.

\begin{figure}
\includegraphics[angle=0,scale=1.4]{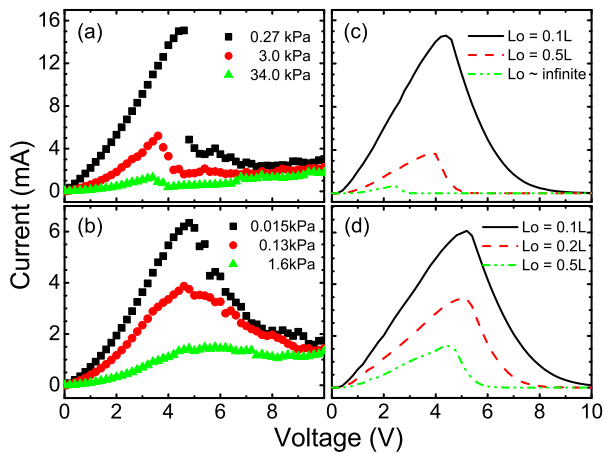}
\caption{Experimental (left panels) and corresponding theoretical
(right panels) results of ON state current dependence on the oxygen
pressure and $L_o$ for different device structures: (a) ITO/Alq$_3$
(120nm)/Au; (b) ITO/Alq$_3$ (120nm)/Al; and different parameters:
(c) $J=0.55$eV, $U_0=1.2$eV; (d) $J=0.35$eV, $U_0=1.4$eV. $L_o\sim$
infinite means the oxygen molecule uniformly distributes in the
organic film.}\label{Lo}
\end{figure}

To verify the present theory, we have to test it for independent
experiments. Here, we show the results for different oxygen
pressures, in Fig. \ref{Lo}. The left panels are for the ON state
current of different structures under various pressures. It shows
that, with increasing pressure, the current reduces very quickly. As
the OFF state current is also much smaller than the ON state, this
reduction of current is quite different from ON-OFF transition. The
reason is that, compared with Fig. \ref{U0}, there's no turn-on
point as the OFF state current shows, and also the downward sides
separate largely. This indicates the oxygen plays a different role
with the interfacial metallic tip, which matters in the ON-OFF
transition. Another important point is that, we have also found (not
shown here) the reduction of current is reversible, namely, when the
pressure of oxygen is reset to a small value, the ON state current
revives. More details of the experiment will be reported elsewhere.
From this experiment, we can see that, the present of oxygen
environment suppresses the memory effect heavily, which implies the
essential role of oxygen in this type of memory devices.

In a common sense, the oxygen penetrates into the organic layer
through the top electrode, due to the loose structure of organic
materials. Hence, the penetration length $L_o$ is intuitively
proportional to the pressure of oxygen. To study this effect more
comprehensively, we could then change the $L_o$ to check if the
theoretical result behaves the same as experiment. As the right
panels of Fig. \ref{Lo} show, we have rebuilt the experimental data,
except within the high bias voltage region, in which the IV curves
of devices undergo a shift due to some unknown cause. Typically,
both the turn-on point and the downward side of the curves are the
same with those from experiment. The explanation of this finding is
as follows. When we change $U_0$, it means the potential drop among
the islands becomes quicker, such that the maximum value shifts to
higher bias voltage. However, when we change $L_o$, it means the
potential drop in the film becomes slower, but the injection at the
bottom electrode becomes harder. Within this understanding, the
maximum point does not change very much, but the whole current curve
goes down. This description could also explain why in those
materials with good conducting performance, such as pentacene and
CuPc, the memory effect is less robust.\cite{You1} That is, in these
materials, the density of electron trapped by oxygen is larger than
normal cases.

%In the last part, we study the dependence of ON state current on
%$J$. $J$ is dependent on the concentration of metallic island, which
%is related to, e.g., the thickness of the organic film. It was found
%that, even if the average electric field is the same, when we change
%the thickness of film, the switch time of the device from OFF to ON
%becomes much larger, and the ON state current becomes
%smaller.\cite{You2} In Fig. \ref{Ji}, the relation between current
%and $J$ is shown. We can find a turn-on point at about $J=0.2$eV,
%and then the current increases quickly till saturation. We expect
%that, these phenomena of both turn-on and saturation could be
%observed in experiment by changing the thickness of organic layer in
%a large region.

Finally, we would like to discuss the other type of organic memory
device, say for example, that based on the electrochemical
reaction.\cite{Island} It was proposed that, a creation-and-rupture
process of redox-controlled metallic bridges is respondence to the
memory effect. Although this electrochemical process is quite
different from the formation of metallic tips, to simulate it one
can still act to change the interfacial potential as we do in this
work. It implies that, the microscopic origin for different types of
organic memory devices might be different, but the underlying
mechanisms are intrinsically the same.

%\begin{figure}
%\includegraphics[angle=0,scale=0.7]{fig4}
%\caption{Dependence of ON state current on $J$. $U_0=1.0$eV,
%$L_o=0.1L$}\label{Ji}
%\end{figure}

In summary, we have presented a theoretical model combining the
effect of tunneling among metallic islands and interfacial electron
emission due to the interfacial metallic tips to comprehensively
investigate the underlying mechanism of organic memory devices. We
calculate several sets of theoretical results, which compare well
with corresponding experimental measurements in different devices.
Based upon this comparison between theory and experiment, we
conclude that the two essential effects in these devices, say memory
effect and negative differential resistance, originate from the
formation of metallic tips at interface and tunneling among metallic
islands in the bulk, respectively. To further prove this conclusion,
we show both theoretical and experimental results on the influence
of oxygen, which shows a clear oxygen pressure dependence. Finally,
we give an expectation on the influence by changing the thickness of
organic film.

The authors would like to acknowledge the financial support from the
National Natural Science Foundation of China and the National Basic
Research Program of China (2012CB921401 and 2009CB929204).

\end{document}